\documentclass{IET-Conf-Paper}

\usepackage{amsmath,amssymb,amsfonts}
\usepackage{xcolor}
\usepackage{algorithm}
\usepackage{algpseudocode}
\usepackage{array}
\usepackage{soul}

\definecolor{bluePlot}{HTML}{0072BD}
\definecolor{yellowPlot}{HTML}{EDB120}
\definecolor{orangePlot}{HTML}{D95319}

\newtheorem{remark}{Remark}[section]

\newtheorem{definition}{Definition}[section]

\begin{document}

\title{DATA-DRIVEN CURRENT CONTROL OF MODULAR MULTILEVEL CONVERTER-BASED HVDC TRANSMISSION SYSTEM}

\author{Victor~D.~Reyes~Dreke\ad{1}\corr, Rahul~Rane\ad{1}, Aleksandra~Leki\'{c}\ad{1}}

\address{\add{1}{ Intelligent Electrical Power Grids (IEPG) group, Delft University of Technology, Delft, The Netherlands}
\email{v.d.reyesdreke@tudelft.nl}}


\begin{abstract}
Model-based design (MBD) is a mature framework for designing the current controllers of modular multilevel converters (MMCs) in high-voltage direct current (HVDC) transmission systems.
In MBD, linear state-space models are preferred for their good trade-off between accurate dynamics prediction and simplicity.
However, first-principle derivations often neglect relevant MMC dynamics, which can hinder controller performance.
Data-driven (DD) methods can deliver less biased models that capture parameter uncertainties, control-loop interactions, and grid dynamics, yet they remain under-exploited for controlling MMC currents.
This paper studies DD current control method based on integral Subspace Predictive Control (iSPC), synthesized directly from input--output data measured in a high fidelity real-time digital simulator (RTDS\textsuperscript{\textregistered}), and proposes an initial guideline outlining the main steps for its systematic application to MMC current control.
The proposed iSPC is compared against state-of-the-art DD solution, and the findings are validated through simulation studies in MATLAB/Simulink.
\end{abstract}

\maketitle

\section{Introduction}
\label{sec:introduction}

Nowadays, high-voltage power transmission is increasingly based on direct current (DC) technology~\cite{Ahmed2011_HVDC_General}.
In high-voltage direct current (HVDC) systems, modular multilevel converters (MMCs) have become the preferred solution for power conversion owing to their high efficiency, scalability, robustness, and low harmonic distortion~\cite{Marquardt2018_MMCandHVDC}.
However, the performance and safe operation of MMCs strongly depend on the design of their control systems, particularly the current controllers~\cite{Beerten2014_HVDC_Control}. 

A mature framework for designing the current controllers of MMC is model-based design (MBD), where mathematical models and control theory are exploited to systematically synthesize controllers that offer desired closed-loop properties, such as stability, and robustness~\cite{Dreke2026_MMC_Control}.  
Among the examples of MBD, we can highlight optimal control~\cite{Ayari2017_MMC_Control}, robust control~\cite{Tavakoli2022_MMC_RControl2}, and model predictive control (MPC)~\cite{Shetgaonkar2023_MMC_MPC}.
Here, linear state-space (SS) models are the preferred model structure due to their good trade-off between accurate dynamics prediction and computational complexity.
These features are particularly important in MPC, where prediction models are used to compute optimized control actions while explicitly considering converter constraints and future system behavior~\cite{Reyes_2022_MMC}.
As a consequence, MPC has demonstrated the ability to enhance MMC performance, including efficiency and power quality, compared with other control methods~\cite{Bocker2014_MMC_MPC}.

Despite these advantages, the effectiveness of MBD strongly depends on the accuracy of the underlying model.
When using first-principle SS (fSS) modeling, models rely on small-signal theory and often neglect practical effects such as digital delays, filtering, and interactions between control layers (including phase-locked loop), which may degrade performance or even induce instability~\cite{Rane2025_MMC_SI}. 
These shortcomings have motivated a shift toward data-driven (DD) modeling, where frequency scanning is often used to generate input--output (I/O) data that allows to build less biased model, see e.g.~\cite{Rane2025_MMC_SI,Gong2018_FREQ_SCAN}. 
These models have been successfully applied to enhance stability and interaction studies; however, these advantages have not yet been leveraged in the design of MMC current controllers.

DD control is a promising alternative to MBD, synthesizing controllers from measured I/O data. 
Prominent examples, including data-driven LQR~\cite{Silva2019_LQRDataDriven}, Subspace Predictive Control (SPC)~\cite{Favoreel1999_SPC}, and Data-Enabled Predictive Control (DeePC)~\cite{Coulson2019_DeePC}, have been applied in power-system and power-electronics applications~\cite{Bilgic2022_DPC_Survey,Leng2026_DPC_Survey,Kraayeveld2027_iSPC_Survey}, yet remain unexplored for MMC current control. 
Their systematic implementation in this setting is still an open problem, as the closed-loop performance hinges on design choices with no counterpart in MBD~\cite{Verheijen2023_handbook_data_driven}, namely the system identification (SI) experiment design (excitation signal, amplitude, and data length), the satisfaction of persistency-of-excitation conditions~\cite{Willems2005_PE}, and the choice of horizons and regularization. 
In practice, they are still selected ad hoc due to the complexity of systematically applying traditional SI procedures~\cite{Ljung1999_SysID}, showing the need for design guidelines.


Hence, this paper proposes using DD, i.e., an integral SPC (iSPC), for current control of MMCs in HVDC systems.
We first design the SI experiment to obtain informative I/O data measured in a high fidelity real-time digital simulator (RTDS\textsuperscript{\textregistered}), which capture the commonly neglected dynamics.
Using these data, the iSPC is designed and compared with DeePC, showing more accurate tracking and lower sensitivity to noise at a lower computation time.
Based on extensive MATLAB/Simulink simulations, we further propose guidelines for tuning the iSPC, covering the selection of the prediction horizon, the data length, and excitation signal type.
\begin{figure*}[htp]
    \centering
    \includegraphics[width=\linewidth]{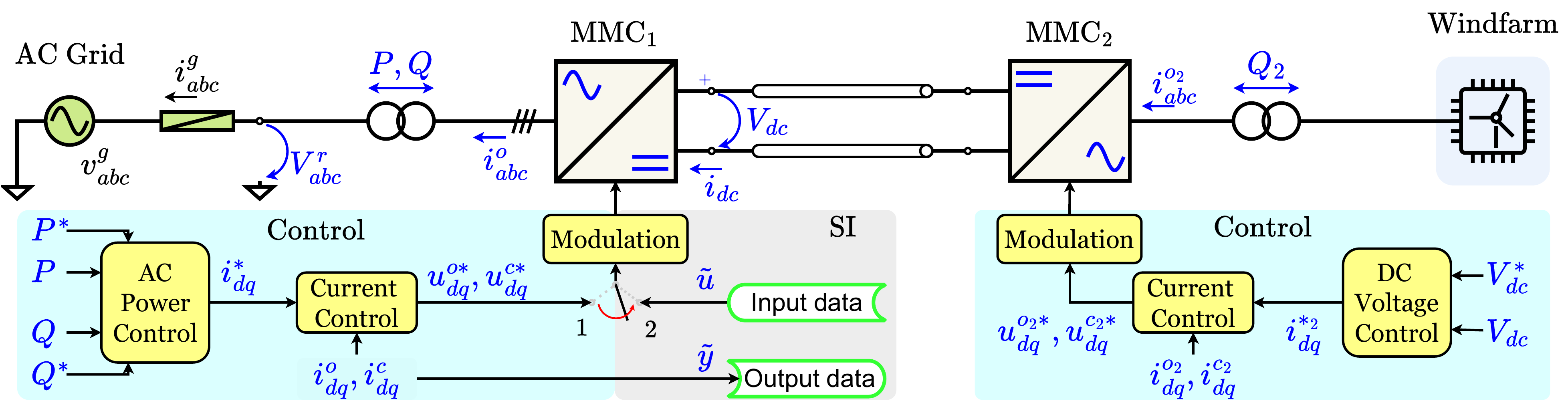}
    \caption{Schema of point-to-point two terminals MMC-based HVDC link including control and SI stages.}
    \label{fig:HVDC_Links_Schemes_Full}
\end{figure*} 
\section{MMC-based HVDC Fundamentals}
\label{sec:mmc_based_hvdc_fundamentals}
Fig.~\ref{fig:HVDC_Links_Schemes_Full} shows the topology of a point-to-point two-terminal MMC-based HVDC link, emulating an offshore wind farm connected to an onshore AC grid.
The onshore MMC$_1$ controls the power flowing to the AC grid, while the offshore MMC$_2$ regulates the DC voltage of the DC link.  
Three-level hierarchical control architectures (HCAs) are often used to control MMC~\cite{Du2018MMCBook}. 
At the highest level, the outer control loop regulates the active/reactive AC power or DC voltage, depending on the MMC operation mode.
This control layer outputs are the current references used in the current control loop (CCL), i.e., $i^*_{dq}{=}(i^{o*}_{dq}, i^{c*}_{dq})$ where the circulating is set to zero $i^{c*}_{dq}{=}0$. 
Then, the CCL computes arm voltage references $(u^{o*}_{dq}, u^{c*}_{dq})$ that regulate MMC currents. 
At the lowest level, $(u^{o*}_{dq}, u^{c*}_{dq})$ are converted into modulation indices and switching signals for the submodule gates. 

Commissioning MMC controllers is often a complex process.
Besides the practical effects, MMC controllers must handle nonlinear dynamics arising from internal arm energy, modulation, and interactions between control layers.
Linear fSS models neglect these dynamics by taking the arm voltages as control inputs and assuming that they equal their references $(u^{o*}_{dq}, u^{c*}_{dq})$~\cite{Du2018MMCBook}.
Hence, controller performance often relies on ad-hoc fine-tuning at later stages.

\subsection{MMC Current Dynamics}
\label{sec:Fundamentals}

\begin{figure}[htp]
    \centering
    \includegraphics[width=\linewidth]{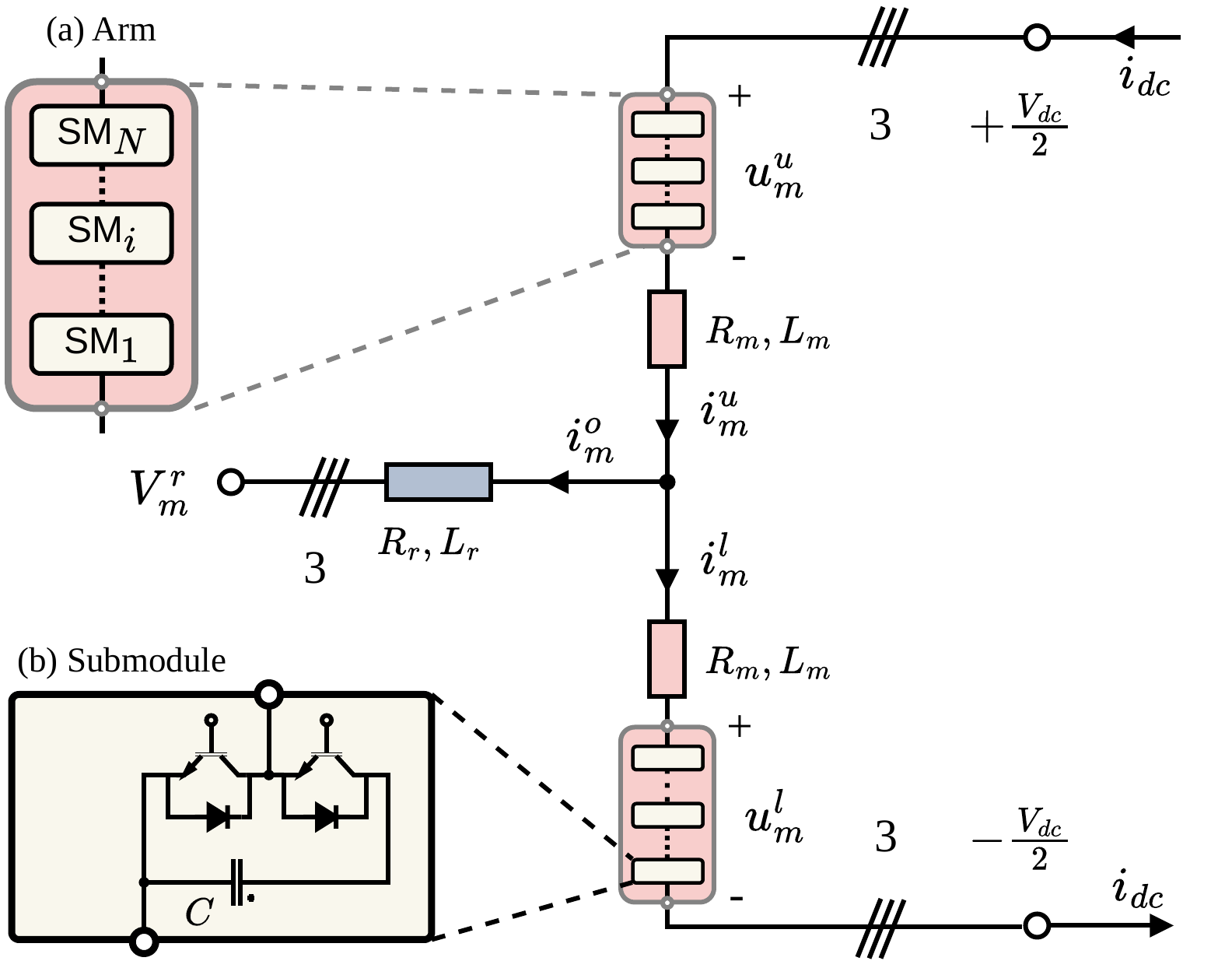}
    \caption{Simplified diagram of the MMC topology per-phase $m \in \{a,b,c\}$: (a) MMC arms; (b) Half-bridge submodule.}
    \label{fig:MMC_SEST26_}
\end{figure}

MMCs are comprised of 3 legs with two arms, in which $N$ submodules  
(SMs) are connected in series to an equivalent resistor ($R_m$) and inductor ($L_m$), see Fig.~\ref{fig:MMC_SEST26_}. 
The arms are named as upper and lower arm, denoted by ${n \in \{u,l\}}$, and the phases are denoted by $m \in \{a,b,c\}$. 
Then, Kirchhoff's voltage law shows that MMC currents are governed by the transformer \textcolor{black}{(TF)} 
$(V^r_{m})$, DC-link $(V_{dc})$, and arm voltages $(u^{u,l}_{m})$, such that:
\begin{subequations}
\label{eq:DifferentialEquationUpper}
    \begin{align}
        \label{eq:DifferentialEquationUpper1}
             L_{m} \frac{\mathrm{d}\, {i}_{m}^{{u}}}{\mathrm{d} t}  &=-R_{m} {i}_{m}^{{u}}+u^u_{m}+V^r_{m}+  L_{r} \frac{\mathrm{d}\, {i}_{m}^{o}}{\mathrm{d} t}+R_{r}{i}_{m}^{o}-\frac{V_{dc}}{2}, \\
             \label{eq:DifferentialEquationUpper2}
             L_{m} \frac{\mathrm{d} \, {i}_{m}^{l}}{\mathrm{d} t}  &=-R_{m} {i}_{m}^{l}+u^l_{m}-V^r_{m}  -L_{r}\frac{\mathrm{d}\, {i}_{m}^{o}}{\mathrm{d} t}-R_{r}{i}_{m}^{o}-\frac{V_{dc}}{2},
    \end{align}
\end{subequations}
where $L_{m}$ and $L_{r}$ are the arm and transformer inductance, respectively; and $R_{m}$ and $R_{r}$ are the arm and transformer resistance, respectively. 
Per phase \( m \), the MMC currents determine the power transfer through the output current, i.e., $i^{o}_{m} = i_m^u - i_m^l$ and circulating current, i.e., $i^{c}_{m} = \tfrac{1}{2} (i_m^u + i_m^l)$. Related to these currents, there are the differential voltage, i.e., $u^{o}_{m} = u_m^u - u_m^l$ and common voltage, i.e., $u^{c}_{m} = \tfrac{1}{2} (u_m^u + u_m^l)$.
\subsection{MMC Model: First-principle state-space derivation}
Assuming an average equivalent representation of the MMC as in \cite{Reyes_2022_MMC} and using the modeling procedures in \cite{Shetgaonkar2023_MMC_MPC}, the dynamics of \( (i^o_m, i^c_m )\) in  \eqref{eq:DifferentialEquationUpper} are formulated in the $dqz$-frame as discrete fSS models\footnotemark{} with sampling time $T_s$ where the state, control input and output vectors are defined as $x := [i^o_d\,,\,i^o_q\,,\,i^c_d\,,\,i^c_q]^{\top}$, ${u := [u^o_d-V^r_d/2\,,\,u^o_q-V^r_q/2\,,\,u^c_d\,,\,u^c_q]^{\top}}$, and $y = x$, respectively. 
\footnotetext{Notation $x^{+}$ describes the next sample, such that $x^{+} \triangleq x(k+1)$. }
Then, its dynamics yield: 
\begin{equation}
    \label{eq:ss_overall_output_current_dq}
    \begin{aligned}
         x^+ &=\textbf{A} x+ \textbf{B}u,
    \end{aligned}       
\end{equation}
with $\textbf{A} {=} e^{\textbf{A}_c{T}_{s}}$ and $\textbf{B}{=} \int_0^{T_s}  e^{\textbf{A}_{c}\tau} \textbf{B}_{c} d\tau$ are state-space matrices, such that $\textbf{A}_c {=} \operatorname{diag}(\textbf{A}^o_c,\textbf{A}^c_c)$ and $\textbf{B}_c {=} \operatorname{diag}(\textbf{B}^o_c,\textbf{B}^c_c)$, where
\begin{equation}
       \begin{aligned}
       \textbf{A}^o_c &= \begin{bmatrix}
        -\frac{R_{eq}}{L_{eq}} & -\omega \\
            \omega & -\frac{R_{eq}}{L_{eq}}
        \end{bmatrix}, \;\;\,   
           \textbf{A}^c_c = \begin{bmatrix}
        -\frac{R_{m}}{L_{m}} & 2\omega \\
            -2\omega & -\frac{R_{m}}{L_{m}}
        \end{bmatrix},
       \end{aligned} 
    \end{equation}
 $\textbf{B}^o_c = \operatorname{diag}{(\frac{1}{L_{eq}},\frac{1}{L_{eq}}  )}$, $\textbf{B}^c_c = \operatorname{diag}{(-\frac{1}{L_{m}},-\frac{1}{L_{m}})}$, $L_{eq} = L_{r}+ \frac{L_{m}}{2},$ $R_{eq} = R_{r}+ \frac{R_{m}}{2}$ and $\omega = 2\pi f$. Operator $\operatorname{diag}(\cdot)$ defines a diagonal matrix where the diagonal elements are its argument.

The augmented model is derived to enforce zero-error tracking when following step references. 
Discrete-time derivation is denoted by the operator $\delta \cdot$, such that ${\delta x(k):= x(k) -x(k-1)}$.
The augmented state vector is defined as $\vec{x} = [\delta {x}^{\top} \; {y}^{\top}]^{\top}$, and an input rate vector, i.e, $\delta u(k) =u(k) -u(k-1)$. 
Then,  \eqref{eq:ss_overall_output_current_dq} augmented model yields
\begin{equation}
\label{eq:ss_output_current_augmented_dqz}
\begin{aligned}
     \vec{x}^{+} &= \vec{\textbf{A}} \vec{x} + \vec{\textbf{B}} \delta u,\quad
     y &= \vec{\textbf{C}} \vec{x},
\end{aligned}   
\end{equation}
where  $\vec{\textbf{A}} {=} \begin{bmatrix}
        \textbf{A} & 0_4\\
        \textbf{C} \textbf{A}  &\textbf{I}_4
    \end{bmatrix} , \; 
        \vec{\textbf{B}}{=} \begin{bmatrix}
        \textbf{B}\\
       \textbf{C} \textbf{B}
    \end{bmatrix},\;
         \vec{\textbf{C}}{=} \begin{bmatrix}
             0_4 & \textbf{I}_4
         \end{bmatrix}$ and $\textbf{C} {=} \textbf{I}_4$.
 Notation $\textbf{I}_4$ indicates an identity matrix of dimension 4, and $0_4$ indicates a square matrix of dimension 4 filled with zeros.
\section{Data-driven predictive control fundamentals}
\label{sec:data_driven_fundamentals}
MBD-MPC computes control actions, which in this case are $\delta u$, via solving at each discrete time step $k$ an optimal control problem (OCP), that can be represented as:
\begin{subequations}
\label{eq:OPC}
    \begin{align}
    \label{eq:OPC_min}
        \min_{\delta U_k}& \;\; J(y(0),\delta U_k)\\
        \label{eq:equality_const}
                    \text{s.t.}& \;\; \vec{x}_{h+1\mid k} = \vec{\textbf{A}} \vec{x}_{h\mid k} + \vec{\textbf{B}} \delta u_{h\mid k}, \; y = \vec{\textbf{C}} \vec{x}_{h\mid k},\\
                    \label{eq:ineq_const}
                                &\;\; y_{\min}\leq y\leq y_{\max},\; u_{\min}\leq u\leq  u_{\max},
    \end{align}
\end{subequations}
with a cost function 
\begin{equation}
\label{eq:costFunctionDef}
    J(\cdot) {=} e^{\top}_{N|k} \; Q \; e_{N|k} +\sum^{N-1}_{h=0} e^{\top}_{h|k} \; Q \; e_{h|k} + \delta{u}^{\top}_{h|k} \; R \; \delta{u}_{h|k},
\end{equation}
where $e_{h|k}{=} y_{h\mid k}{-}r(k+h)$ is the predicted error\footnotemark{}, $N$ is the prediction horizon, $(Q,R)$ are weighting matrices, $(\vec{x}_{\min},\vec{x}_{\max})$ are state limits,  $( u_{\min}, u_{\max})$ are input limits. 
\footnotetext{Notations $e_{h|k}$, $\delta{u}_{h|k}$, ${y}_{h|k}$ and $\vec{x}_{h\mid k}$ refers to $h$-ahead predictions, e.g., $e_{h|k} = y(h+k){-}r(h+k)$ and $e_{0|k} = y(k){-}r(k)$.}

DD predictive control (DPC) follows the MBD-MPC principle, except that the prediction step, i.e., \eqref{eq:equality_const}, relies on an identified model or directly on measured I/O data rather than on the fSS.
The I/O data are collected through an SI experiment, in which an excitation signal applied to the arm voltages forms the input sequence $\tilde{u}$ and the measured currents $(i^o_{dq},i^c_{dq})$ form the output sequence $\tilde{y}$ (see Fig.~\ref{fig:HVDC_Links_Schemes_Full}).
DPC builds on two key concepts: Hankel matrix and persistence of excitation.
A Hankel matrix arranges a data sequence  $(\tilde{y},\tilde{u})$ in a fixed block structure; following \cite{Verheijen2023_handbook_data_driven}, it is defined as
\begin{definition}
\label{def:Hankel_Matrix}
    A Hankel matrix constructed using an arbitrary dataset $z$, with depth $N$, length $T$ and starting index $i$ is denoted as:
    \begin{equation*}
    \resizebox{\columnwidth}{!}{%
        $\mathcal{H}_{[N, T]}^i(z){=}\left[\begin{array}{cccc}
        z(i) & z(i+1) & \ldots & z(i+T-1) \\
        z(i+1) & z(i+2) & \ldots & z(i+T) \\
        \vdots & \vdots & & \vdots \\
        z(i+N-1) & z(i+N) & \ldots & z(i+T+N-2)
        \end{array}\right]$%
        }
    \end{equation*}
    where the samples in $z$ can comprise information incoming from multiple channels (e.g., multiple inputs or outputs), i.e., $z(i)=$ $\left[\begin{array}{llll}z_0(i) & z_1(i) & \ldots & z_r(i)\end{array}\right]^T$.
\end{definition}
The I/O data collected during SI, i.e., $(\tilde{y},\tilde{u})$, are used to construct the following Hankel matrices:
\begin{equation}
\begin{aligned}
    \mathbf{U}_p&=\mathcal{H}_{\left[T_{\text {ini }}, T\right]}^0(\tilde{u}), \quad  \mathbf{Y}_p=\mathcal{H}_{\left[T_{\text {ini }}, T\right]}^1(\tilde{y}),  \\
\mathbf{U}_f&=\mathcal{H}_{[N, T]}^{T_{\text {ini }}}(\tilde{u}), \quad \;\,\mathbf{Y}_f=\mathcal{H}_{[N, T]}^{T_{\text {ini }}+1}(\tilde{y}),
\end{aligned}
\end{equation}
where $T_{\text{ini}}$ is the "past horizon".
These matrices are at the core of DPCs, namely SPC, DeePC, $\gamma$-DDPC and GDPC \cite{Verheijen2023_handbook_data_driven}.
\begin{remark}
    The past horizon $T_{\text{ini}}$ is related to the system order $n$ and should satisfy $T_{\text{ini}} \geq n$; for the current dynamics, $n = 8$, hence $T_{\text{ini}} \geq 8$.
    To ensure that the data are \emph{informative}, the Hankel matrix length must satisfy $T \geq n_u\left(T_{\text{ini}} + N + n\right)$, where $n_u = 4$ is the input dimension.
    Then, we define the data length as $T_{\text{data}}=T_{\text{ini}}+T+N$, such that 
    $(\tilde{y},\tilde{u}) = \{(\tilde{y}(0),\tilde{u}(0)), {...}, (\tilde{y}(T_{\text{data}}),\tilde{u}(T_{\text{data}}))\}$.
\end{remark}
Additionally,  informativity  can be ensured if $\tilde{u}$ is persistently exciting, defined as  
\begin{definition}[Persistence of Excitation]
    The input data sequence $\tilde{u}$ is called persistently exciting of order $L$ if the rank of the Hankel matrix satisfies
    \begin{equation}
        \operatorname{rank}\left(\mathcal{H}_{[L, T]}^0(\tilde{u})\right)=L n_u 
    \end{equation}
    where $n_u$ is the dimension of the control inputs. 
\end{definition}
\subsection{Integral Subspace Predictive Control (iSPC)}
iSPC is an indirect DPC~\cite{Kraayeveld2027_iSPC_Survey} that estimates, from I/O data, a prediction matrix $\Theta$ that computes $N$-step-ahead, such that
\begin{equation}
\label{eq:prediction_model_spc}
   Y_k=\underbrace{\left[\begin{array}{lll}
    P_1 & P_2 & \Gamma
    \end{array}\right]}_{:=\Theta}\left[\begin{array}{c}
    \mathbf{u}_{\mathrm{ini}}(k) \\
    \mathbf{y}_{\mathrm{ini}}(k) \\
    \delta U_k
    \end{array}\right], 
\end{equation}
where $Y_k {=} [y^{\top}_{1\mid k},...,y^{\top}_{N\mid k}]^{\top}$, $\delta U_k{=}[\delta u^{\top}_{0\mid k},...,\delta u^{\top}_{N{-}1\mid k}]^{\top}$, $\mathbf{u}_{\mathrm{ini}}(k) = \left[\delta u^{\top}(k-T_{\text{ini}}),...,\delta u^{\top}(k-1) \right]^{\top}$ and $ \mathbf{y}_{\mathrm{ini}}(k) = \left[ y^{\top}(k-T_{\text{ini}}),...,\right.$ $\left.y^{\top}(k-1) \right]$.
According to \eqref{eq:prediction_model_spc} future outputs $Y_k$ are described as a linear combination of previous I/O data and $\delta U_k$.
By relying on this relation, the estimation of the unknown matrix $\Theta$ can be carried out by solving the least-squares problem:
\begin{equation}
    \min _{\Theta}\left\|\mathbf{Y}_f-\Theta\left[\begin{array}{l}
        \mathbf{U}_p \\
        \mathbf{Y}_p \\
        \mathbf{U}_f
        \end{array}\right]\right\|_{\text {Fro }}^2,
\end{equation}
with solution $\Theta^*$, i.e.,
\begin{equation}
\label{eq:estimatedTheta}
   \Theta^*=\left[\begin{array}{lll}
    P_1 & P_2 & \hat{\Gamma}
    \end{array}\right]=\mathbf{Y}_f\left[\begin{array}{l}
    \mathbf{U}_p \\
    \mathbf{Y}_p \\
    \mathbf{U}_f
    \end{array}\right]^{\dagger},
\end{equation}
where $\hat{\Gamma}$ is the DD estimation of $\Gamma$ and $^{\dagger}$ indicates the pseudo-inverse matrix. 
Using \eqref{eq:estimatedTheta}, iSPC yields
\begin{subequations}
\label{eq:SPC_MPC}
    \begin{align}
    \label{eq:OPC_SPC_min}
        \min_{\delta U_k}& \;\; J(y(0),\delta U_{k})\\
        \label{eq:SPC_equality_const}
                    \text{s.t.}& \;\; Y_k=P_1 \mathbf{u}_{\mathrm{ini}}(k)+P_2 \mathbf{y}_{\mathrm{ini}}(k)+\hat{\Gamma} \delta U_k,\\
                    \label{eq:SPC_ineq_const}
                                &\;\; y_{\min}\leq y\leq y_{\max}, \;u_{\min}\leq u\leq  u_{\max},
    \end{align}
\end{subequations}
where the cost function $J(y(0),\delta U_{k})$ stays as in \eqref{eq:costFunctionDef}.
The iSPC implementation is summarized by the following algorithm. 
\begin{algorithm}
\caption{SPC algorithm}
\begin{algorithmic}[0]
\Require $\mathbf{U}_p, \mathbf{Y}_p, \mathbf{U}_f, \mathbf{Y}_f, (N, T_{\text{ini}}, Q, R, r)$
\Statex \textbf{Before closing the control loop:}
\State 1: Obtain $\Theta^*$ from \eqref{eq:estimatedTheta} and extract $P_1$, $P_2$ and $\hat{\Gamma}$.
\State 2: Construct controller as in \eqref{eq:SPC_MPC}.
\Statex \textbf{During the control loop:}
\State 1: Measure $y(k)$ and construct $\mathbf{u}_{\text{ini}}(k)$ and $\mathbf{y}_{\text{ini}}(k)$.
\State 2: Solve control law \eqref{eq:SPC_MPC} and obtain $\delta U^*_{k}$.
\State 3: Apply $u(k) = u(k-1)+\delta u^*_{0|k}$ to the system.
\end{algorithmic}
\end{algorithm}
\section{Case Study}
\label{sec:Case_Study}
The iSPC performance is studied in a point-to-point, two-terminal MMC-based HVDC link, i.e., CIGRE DCS1 RTDS\textsuperscript{\textregistered} benchmark \cite{CIGRE_TB604_2014}, rated at 0.8~GVA and 50~Hz, with a DC-link voltage of 400~kV.
On the AC side, both MMCs operate at 220~kV and are interfaced through TFs to the 380~kV onshore and 145~kV offshore grids; the TFs have a resistance of 0.006~pu and an inductance of 0.18~pu.
Each MMCs, modeled with a type 5 average value model~\cite{CIGRE_TB604_2014}, have 200 SMs per arm with a 10~mF capacitance, $R_m{=}0.15$~$\Omega$ and $L_m{=}29$~mH.
CIGRE DCS1 benchmark~\cite{CIGRE_TB604_2014} is a high fidelity model that includes SM-level dynamics, practical effects, and a three-level HCA, detailed in~\cite[Ch.~3]{CIGRE_TB604_2014}. iSPC is proposed to replace only the CCL PI controllers of the control architecture.
\subsection{ System identification and validation}
\begin{figure}[htp]
    \centering
    \includegraphics[width=\linewidth]{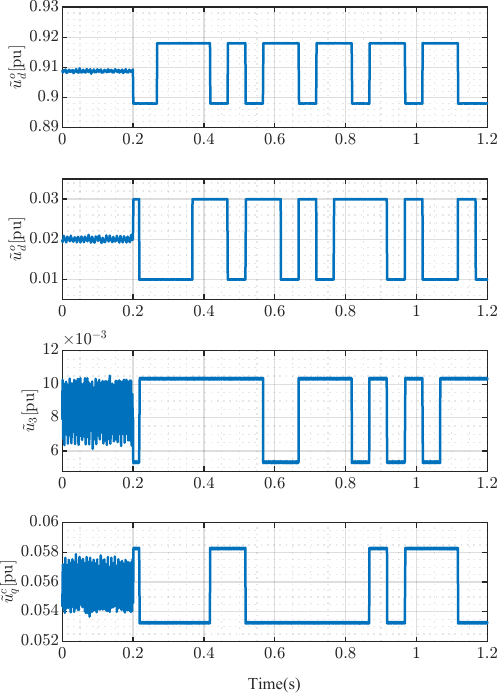}
    \caption{Input data sequence, i.e., $\tilde{u}(t) =\{\tilde{u}^o_d,\tilde{u}^o_q,\tilde{u}^c_d,\tilde{u}^c_q, \}$ .}
    \label{fig:figureIdSysInput}
\end{figure}

\begin{figure}[htp!]
    \centering
    \includegraphics[width=\linewidth]{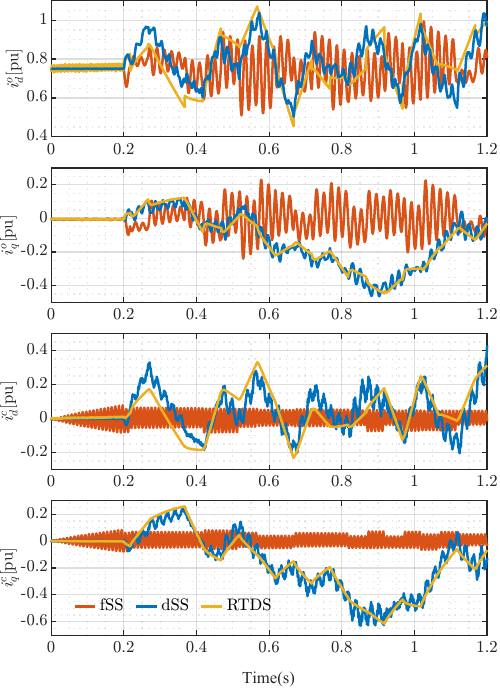}
    \caption{MMC currents generated by $\tilde{u}$ in: (\textcolor{bluePlot}{\textbf{--}}) RTDS , (\textcolor{orange}{\textbf{--}})  fSS, and (\textcolor{yellowPlot}{\textbf{--}}) dSS.}
    \label{fig:figureIdSysOutput}
\end{figure}

\begin{figure}[h!]
    \centering
    \includegraphics[width=\linewidth]{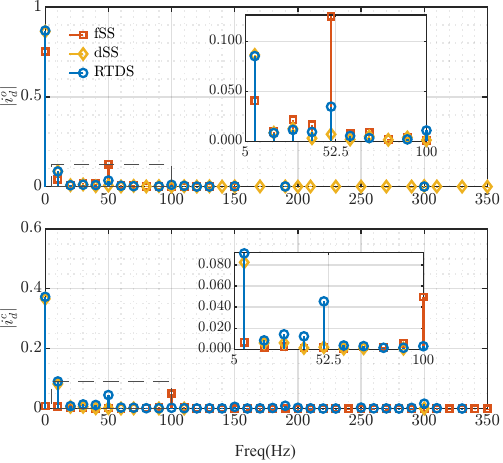}
    \caption{FFT analysis of $i^o_d$ and $i^c_q$ currents in Fig.~\ref{fig:figureIdSysOutput}.}
    \label{fig:figureIdSysOutputFFT}
\end{figure}

Fig.~\ref{fig:figureIdSysInput} presents the excitation input data $\tilde{u}$, i.e, a random binary sequence (RBS) superimposed on white noise:
\begin{equation}
    \tilde{u}_j(k) =
    \begin{cases}
        -\alpha + \varepsilon_j(k), & \text{if } \eta_j(k) \leq 0.5,\\
        \phantom{-}\alpha + \varepsilon_j(k), & \text{if } \eta_j(k) > 0.5,
    \end{cases}
    \qquad j \in \{1,\dots,4\},
\end{equation}
where $\alpha$ is the RBS amplitude, $\eta_j(k) \sim \mathcal{U}(0,1)$ is a uniformly distributed random number whose seed depends on channel $j$, and $\varepsilon_j(k) \sim \mathcal{N}(0,\sigma^2)$ is zero-mean white noise.
Whereas multisine signals are common in the literature, e.g., for frequency scans, the RBS concentrates its energy at low frequencies and captures the steady-state behavior, while the added white noise reduces the correlation between samples and improves the data PE.
Choosing an $\alpha$ large enough to excite the system without violating operating constraints or driving it toward instability is challenging.
We recommend $\alpha$ between 2\% and 6\% of the steady-state value; larger submodule capacitance and arm resistance allow amplitudes toward the upper end of this range. 
For $\varepsilon_j$, a small variance on the order of $\operatorname{var}(\varepsilon_j)\approx10^{-5}$ is suggested to preserve identification accuracy.    
Fig.~\ref{fig:figureIdSysOutput} compares the current measured in RTDS ($f_s=50$~kHz) with the predictions of the fSS model and of a data-driven SS (dSS) model obtained by subspace identification (Appendix~A).
The measurements and dSS matrices reveal dynamics neglected by the fSS model: cross-coupling between the output and circulating current loops, switching and modulation ripple, and pre-processing effects.
The dSS model closely tracks the measured current, particularly its low-frequency content, as confirmed by its lower mean squared prediction error (MSE$_\text{p}$) in Table~\ref{tab:mse}.
The current spectra in Fig.~\ref{fig:figureIdSysOutputFFT} show this more clearly: the fSS model underestimates the 0~Hz component, which sets the gain of the current dynamics, and overestimates the 50~Hz and 100~Hz components relative to RTDS and dSS.
Higher-order linear models could reproduce better the RTDS time response, but would likely overfit rather than yield an unbiased predictor.
These mismatches do not necessarily compromise stability or feasibility, as fSS models have been used effectively in MBD~\cite{Shetgaonkar2023_MMC_MPC}; nevertheless, they reveal a modelling gap that dSS models can close, notably for controller design with theoretical guarantees.
\begin{table}[h!]
\caption{MSEp comparison of  fSS and dSS models}
\begin{center}
\begin{tabular}{lllll}\toprule
MSEp $[\text{pu}^2]$ & $i^o_d$ & $i^o_q$ & $i^c_d$ & $i^c_q$ \\\midrule
fSS & 0.0723 & 0.0828 & 0.0784 & 0.1410 \\
dSS & 0.0044 & 0.0010 & 0.0049 & 0.0024\\\botrule
\end{tabular}
\label{tab:mse}
\end{center}\vspace*{-18pt}
\end{table}
\subsection{iSPC Design and Verification}
\label{subsec:iSPCDesignandVerification}
\begin{figure}[htp]
    \centering
    \includegraphics[width=\linewidth]{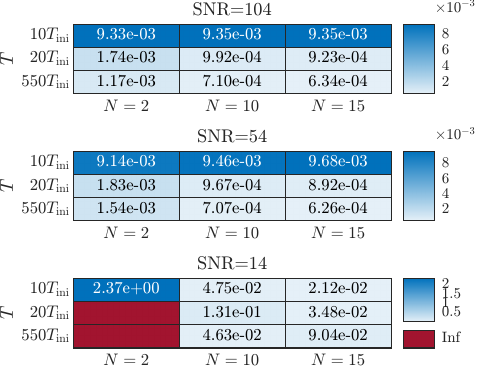}
    \caption{Comparison  of the iSPC tracking performance, i.e., MAE$_\text{t}$, with $Q = \textbf{I}_4$, $R = 10\textbf{I}_4$, and $T_{\text{ini}} =20$ under different conditions of SNR, $T$ and $N$.}
    \label{fig:figureMAESPC}
\end{figure}

Before its implementation in the RTDS, the iSPC is designed and verified in MATLAB.
The weighting matrices are chosen empirically as $Q = \mathbf{I}_4$ and $R = 10\,\mathbf{I}_4$, and the outputs $y = (i^o_{dq}, i^c_{dq})$ and inputs $u = (u^{o}_{dq}, u^{c}_{dq})$ are constrained to $-1.5 \leq y \leq 1.5$~pu and $-1.5 \leq u \leq 1.5$~pu, respectively.
During the design, we examine how $T$ and $N$ affect the trade-off between robustness, control performance, and computational burden, with $T_{\text{ini}} = 20$ fixed.
Robustness is assessed by adding output measurement noise at different signal-to-noise ratios (SNRs), i.e., $\tilde{y}(k) = y(k) + v(k)$, where $v(k)$ is zero-mean white noise.
Fig.~\ref{fig:figureMAESPC} shows the resulting mean absolute tracking error (MAE$_\text{t}$) to step response of the CCL for different SNR, $T$, and $N$.

\begin{figure}[htp]
    \centering
    \includegraphics[width=\linewidth]{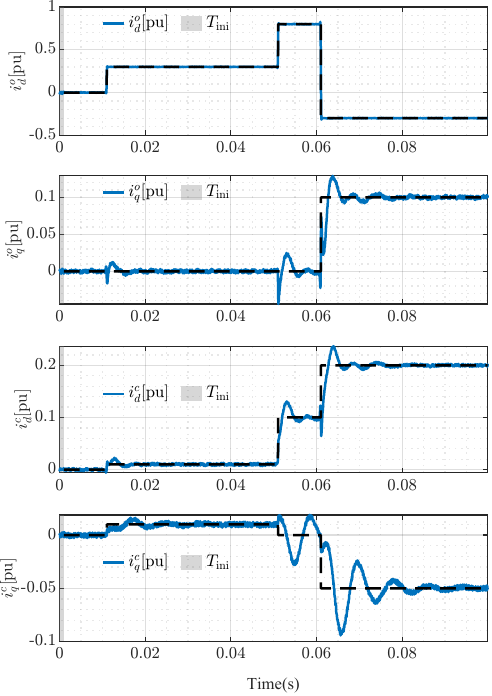}
    \caption{MMC currents under iSPC}
    \label{fig:figureOutputSPC}
\end{figure}

\begin{figure}[htp]
    \centering
    \includegraphics[width=\linewidth]{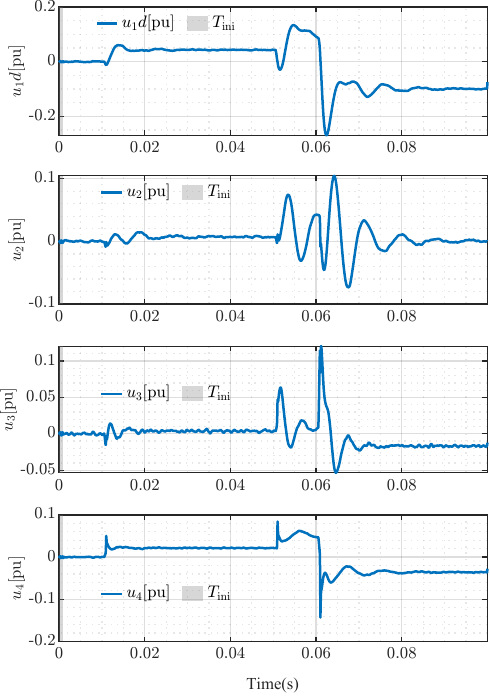}
    \caption{CCL Control input under iSPC}
    \label{fig:figureInputSPC}
\end{figure}

For SNR${=}12$~dB, the added noise $v(k)$ has a variance of $\operatorname{var}(v) {\approx} 0.1022$ and a peak amplitude of $\max|v| = 1.363$~pu.
Under this \emph{highly} noisy condition, the iSPC performance degrades considerably, the MAE$_{\text{t}}$ increases by up to a factor of 100 and even yielding infeasible OCP for $N{=}2$.
Increasing $T$ beyond $20\,T_{\text{ini}}$ yields no substantial improvement under high noise.
Instead, increasing $N$ hold by $N < T_{\text{ini}}$ proves to be the most effective way to improve iSPC performance at a lower computational burden.
The same holds under cleaner conditions (SNR~$=52$~dB and $102$~dB), where increasing $N$ remains the best strategy, whereas for short prediction horizons a larger $T$ does not guarantee better performance.
Finally, increasing $T_{\text{ini}}$, and consequently $T$, yields the best (stable/feasible) performance even under very noisy conditions (SNR$\leq 10$), but at a considerably higher computational cost.
Figs.~\ref{fig:figureOutputSPC} and~\ref{fig:figureInputSPC} show the CCL response to successive reference steps under iSPC in a low-noise scenario (SNR~$\approx52$~dB), the references are tracked without constraint violation.
\subsection{Comparison DeePC vs iSPC}
\begin{figure}[htp]
    \centering
    \includegraphics[width=\linewidth]{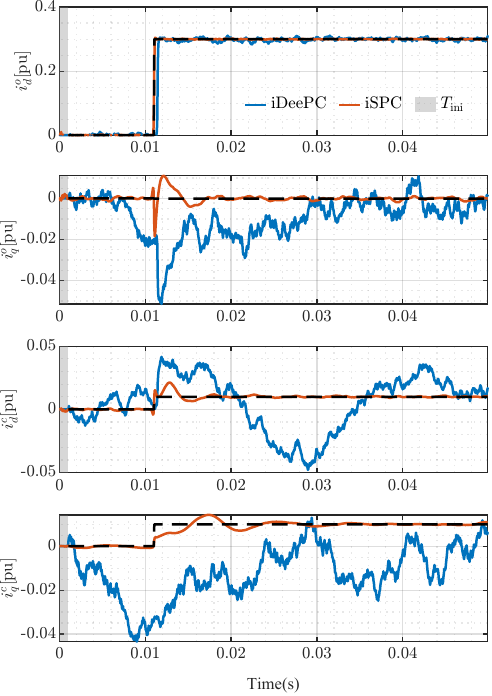}
    \caption{MMC currents under (\textcolor{orangePlot}{\textbf{---}}) iSPC and (\textcolor{bluePlot}{\textbf{---}}) iDeePC.}
    \label{fig:figureOutputSPCvsDeePC}
\end{figure}

\begin{figure}[htp]
    \centering
    \includegraphics[width=\linewidth]{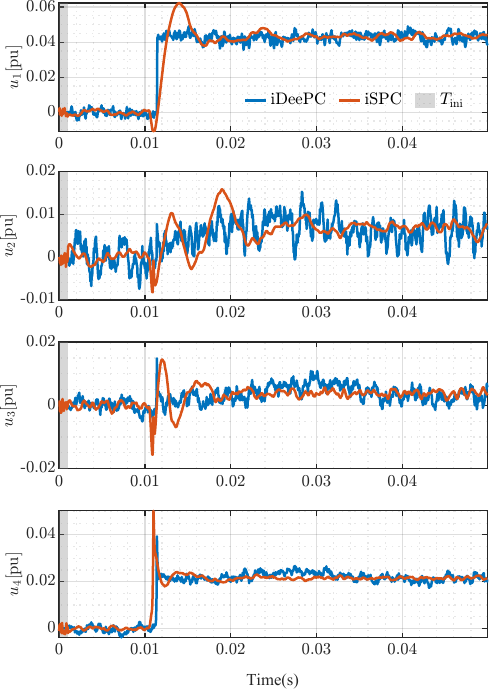}
    \caption{CCL Control input under (\textcolor{orangePlot}{\textbf{---}}) iSPC and (\textcolor{bluePlot}{\textbf{---}}) iDeePC.}
    \label{fig:figureInputSPCvsDeePC}
\end{figure}

We compare the iSPC against DeePC~\cite{Coulson2019_DeePC}, a DPC method widely adopted in the DD control community.
Unlike iSPC, DeePC is a direct method that solves the OCP using the data directly, without an explicit identification step; implementation details can be found in~\cite{Coulson2019_DeePC,Verheijen2023_handbook_data_driven}.
For a fair comparison, both DPCs use the same parameters as the previous section.
For DeePC, the slack variable on the $\mathbf{y}_{\text{ini}}(k)$ is penalized with $\lambda_y = 10^4$, and no regularization on $g$ is applied.
Regularized variants were tested but yielded no significant improvement under the low to moderate noise conditions considered here, while considerably increasing the computation time; hence, they are not reported.
Fig.~\ref{fig:figureOutputSPCvsDeePC} and~\ref{fig:figureInputSPCvsDeePC} show CCL response under iSPC and DeePC with SNR${=}102$~dB, $T_{\text{ini}} {=} 20$, and $T {=} 20\,T_{\text{ini}}$, where the same step references as in Section~\ref{subsec:iSPCDesignandVerification} is considered.
Both controllers track the reference within the constraints; with identical tuning, the iSPC responds faster but with a higher overshoot, whereas DeePC is slower and more sensitive to output noise.
In terms of MAE$_{\text{t}}$, the iSPC achieves $1.82 \times 10^{-5}$~pu, while iDeePC has a $2.29 \times 10^{-4}$~pu.
The integral of the absolute control input (IAU) is $7.7846$~pu for the iSPC and $6.3353$~pu for DeePC.
In MATLAB environment, the average computation time\footnotemark{} per control step is $1.15$~ms for the iSPC and $268$~ms for DeePC. 
\footnotetext{Hardware: Dell Latitude 7340, 13th Gen Intel(R) Core(TM) i7-1365U (1.80 GHz), 16.0 GB RAM. Software: MATLAB 2024b, Mosek solver~\cite{mosek}.}

The iSPC computation time in MATLAB exceeds the $50~\mu$s required for real-time implementation, mainly because the OCP was solved via YALMIP~\cite{lofberg2004yalmip}, a general-purpose modeling toolbox that is not optimised for speed. However, the iSPC structure allows the OCP to be solved with Hildreth's algorithm~\cite{wang2009model} coded in C++, which has proven effective for MMC control~\cite{Shetgaonkar2023_MMC_MPC}. Future work will therefore implement the iSPC directly on RTDS.
\section{Conclusions}
This paper investigated the use of iSPC for current control of MMCs in HVDC systems.
The SI experiment revealed  notable cross-coupling between the output and circulating currents dynamics, often neglected in fSS models.
Consequently, dSS model identified from I/O data were proposed to predict the current dynamics more accurately.
Using these I/O data, we designed the iSPC and proposed a guideline for its systematic application, covering the design of the excitation signal, the selection of $T_{\text{ini}}$ and $T$, and the tuning of the prediction horizon.
Apart from data quality, the prediction horizon $N$ proved to be the most influential design parameter: increasing $N$ yielded the best trade-off between control performance and computational burden, whereas increasing $T$ beyond $20\,T_{\text{ini}}$ brought no substantial improvement in low to moderate noise conditions.
Compared with DeePC under identical tuning, the iSPC achieved a roughly 13 times lower tracking MSE at the cost of a moderately higher control effort, while reducing the computation time by more than two orders of magnitude. Future work will therefore implement the iSPC directly on RTDS.
\section{Acknowledgements}
The work is supported by Horizon Europe project PROSECCO, under grant agreement 101160687, and by NWO Veni project SAFE-GRID, with project number 20248.
\section{Appendix}
fSS matrices derived from CIGRE B4.57 DCS1 benchmark: 
\begin{equation}
A= \begin{bmatrix}  
0.9995 & -0.0157 & 0 & 0 \\
0.0157 & 0.9995 & 0 & 0 \\
0 & 0 & 0.9995 & 0.0314 \\
0 & 0 & -0.0314 & 0.9995 
\end{bmatrix},
\end{equation}
\begin{equation}
B= \begin{bmatrix}  
0.0615 & -0.0005 & 0 & 0 \\
0.0005 & 0.0615 & 0 & 0 \\
0 & 0 & -0.1042 & -0.0016 \\
0 & 0 & 0.0016 & -0.1042
\end{bmatrix}.
\end{equation}
d from CIGRE B4.57 DCS1 benchmark:  
\begin{equation}
A= \begin{bmatrix}  
0.6745 & 0.0615 & 0.3082 & -0.0310 \\
0.0396 & 0.9825 & -0.0368 & 0.0108 \\
-0.0255 & 0.0104 & 1.0237 & -0.0065 \\
-0.0047 & 0.0036 & 0.0044 & 0.9975 \\
\end{bmatrix},
\end{equation}
\begin{equation}
B= \begin{bmatrix}  
0.1961 & 0.9437 & 1.4460 & 3.4511 \\
-0.0565 & -0.1211 & -0.2036 & -0.3503 \\
0.0170 & 0.0774 & 0.1702 & 0.2583 \\
0.0061 & -0.0045 & 0.0230 & 0.0496 \\
\end{bmatrix}.
\end{equation}
\section{References}
\bibliographystyle{IEEEtran}
\bibliography{bibtex/Biblio,bibtex/bib/IEEEabrv,bibtex/bib/IEEEexample}

\end{document}